\documentclass[11pt,onecolumn]{IEEEtran}

\usepackage[mathscr]{eucal}
\usepackage[cmex10]{amsmath}
\usepackage{epsfig,epsf,psfrag}
\usepackage{amssymb,amsmath,amsthm,amsfonts,latexsym}
\usepackage{amsmath,graphicx,bm,xcolor,url}
\usepackage[caption=false]{subfig} 
\usepackage{fixltx2e}
\usepackage{array}
\usepackage{verbatim}
\usepackage{bm}
\usepackage{algorithmic, cite}
\usepackage{algorithm}
\usepackage{verbatim}
\usepackage{textcomp}
\usepackage{mathrsfs}
\usepackage{epstopdf}

\catcode`~=11 \def\UrlSpecials{\do\~{\kern -.15em\lower .7ex\hbox{~}\kern .04em}} \catcode`~=13 

\allowdisplaybreaks[1]

\newcommand{\calA}{\mathcal{A}}
\newcommand{\calB}{\mathcal{B}}
\newcommand{\calC}{\mathcal{C}}

\newcommand{\calE}{\mathcal{E}}
\newcommand{\calF}{\mathcal{F}}

\newcommand{\calH}{\mathcal{H}}

\newcommand{\calJ}{\mathcal{J}}

\newcommand{\calP}{\mathcal{P}}

\newcommand{\calT}{\mathcal{T}}
\newcommand{\calU}{\mathcal{U}}

\newcommand{\calX}{\mathcal{X}}
\newcommand{\calY}{\mathcal{Y}}

\newcommand{\rmd}{\mathrm{d}}

\newcommand{\rme}{\mathrm{e}}

\newcommand{\bbE}{\mathbb{E}}

\newcommand{\bbN}{\mathbb{N}}

\newcommand{\bbR}{\mathbb{R}}

\DeclareMathAlphabet{\mathbsf}{OT1}{cmss}{bx}{n}
\DeclareMathAlphabet{\mathssf}{OT1}{cmss}{m}{sl}

\newcommand{\rvQ}{\mathsf{Q}}

\DeclareSymbolFont{bsfletters}{OT1}{cmss}{bx}{n}  
\DeclareSymbolFont{ssfletters}{OT1}{cmss}{m}{n}
\DeclareMathSymbol{\bsfGamma}{0}{bsfletters}{'000}
\DeclareMathSymbol{\ssfGamma}{0}{ssfletters}{'000}
\DeclareMathSymbol{\bsfDelta}{0}{bsfletters}{'001}
\DeclareMathSymbol{\ssfDelta}{0}{ssfletters}{'001}
\DeclareMathSymbol{\bsfTheta}{0}{bsfletters}{'002}
\DeclareMathSymbol{\ssfTheta}{0}{ssfletters}{'002}
\DeclareMathSymbol{\bsfLambda}{0}{bsfletters}{'003}
\DeclareMathSymbol{\ssfLambda}{0}{ssfletters}{'003}
\DeclareMathSymbol{\bsfXi}{0}{bsfletters}{'004}
\DeclareMathSymbol{\ssfXi}{0}{ssfletters}{'004}
\DeclareMathSymbol{\bsfPi}{0}{bsfletters}{'005}
\DeclareMathSymbol{\ssfPi}{0}{ssfletters}{'005}
\DeclareMathSymbol{\bsfSigma}{0}{bsfletters}{'006}
\DeclareMathSymbol{\ssfSigma}{0}{ssfletters}{'006}
\DeclareMathSymbol{\bsfUpsilon}{0}{bsfletters}{'007}
\DeclareMathSymbol{\ssfUpsilon}{0}{ssfletters}{'007}
\DeclareMathSymbol{\bsfPhi}{0}{bsfletters}{'010}
\DeclareMathSymbol{\ssfPhi}{0}{ssfletters}{'010}
\DeclareMathSymbol{\bsfPsi}{0}{bsfletters}{'011}
\DeclareMathSymbol{\ssfPsi}{0}{ssfletters}{'011}
\DeclareMathSymbol{\bsfOmega}{0}{bsfletters}{'012}
\DeclareMathSymbol{\ssfOmega}{0}{ssfletters}{'012}

\newcommand{\tilD}{\tilde{D}}

\newcommand{\hatP}{\hat{P}}

\newcommand{\hatQ}{\hat{Q}}

\newcommand{\barD}{\bar{D}}

\newcommand{\barU}{\bar{U}}

\newcommand{\eps}{\varepsilon}

\def\fndot{\, \cdot \,}

\DeclareMathOperator{\supp}{supp}

\newtheorem{theorem}{Theorem} 
\newtheorem{lemma}[theorem]{Lemma}

\newtheorem{proposition}[theorem]{Proposition}

\newtheorem{definition}{Definition}

\newcommand{\qednew}{\nobreak \ifvmode \relax \else
      \ifdim\lastskip<1.5em \hskip-\lastskip
      \hskip1.5em plus0em minus0.5em \fi \nobreak
      \vrule height0.75em width0.5em depth0.25em\fi}

\newcommand{\zr}{\mathrm{zr}}

\newcommand{\qyx}{W}
\usepackage{epsfig} 
\usepackage{epstopdf,bbm} 
\usepackage{hyperref}
\usepackage{enumitem}
\usepackage{overpic}
\usepackage{booktabs, tabularx, array}
\usepackage{graphicx}
\usepackage{tikz}
\usetikzlibrary{arrows.meta,patterns}

\usepackage{pgfplots}
\usetikzlibrary{arrows.meta}
\pgfplotsset{compat=1.18}
\usetikzlibrary{arrows.meta,decorations.pathreplacing,shapes.geometric}

\hypersetup{hidelinks}

\usepackage{amsmath,amssymb}
\usepackage{xcolor}
\usetikzlibrary{arrows.meta,calc,positioning,decorations.pathreplacing}

\definecolor{lowerr}{HTML}{D9EEF5}
\definecolor{miderr}{HTML}{F8D49A}
\definecolor{higherr}{HTML}{D96B5F}
\definecolor{keepgreen}{HTML}{2F7F62}
\definecolor{trimgray}{HTML}{D8D8D8}
\definecolor{textgray}{HTML}{555555}

\tikzset{
  stagearrow/.style={-{Latex[length=3.2mm,width=2.2mm]}, very thick, draw=textgray},
  paneltitle/.style={font=\bfseries\small, align=center},
  annotation/.style={font=\scriptsize, align=center, text width=4.35cm},
  axislabel/.style={font=\scriptsize, text=textgray},
  matrixoutline/.style={draw=black!75, line width=0.55pt},
}

\begin{document}
\flushbottom
\title{Third-Order Asymptotics for   Fixed-Length Lossy Compression of Discrete Memoryless Sources} 

\author{Vincent Y.~F.~Tan  %
\thanks{V.~Y.~F.~Tan is with the Department of Mathematics and 
the Department of Electrical and Computer Engineering, National University of Singapore
(email:\,vtan@nus.edu.sg).
}}
 
\IEEEpeerreviewmaketitle

\maketitle

\begin{abstract}
 This paper determines the third-order asymptotics of the logarithm of the minimum codebook size for fixed-length lossy compression of discrete memoryless sources. For an excess-distortion probability constraint $\eps\in(0,1)$ and every interior distortion level, we establish a third-order correction of $\frac12\log n+O(1)$, where $n$ is the blocklength. This sharpens the general bounds of Kostina and Verd\'u
[\emph{IEEE Trans.\ Inf.\ Theory}, 2012].  The main technical innovation lies in the converse for positive dispersion. Under a reverse change of measure, we retain the exponential weights associated with tilted information and distortion slack and control them jointly using the multivariate weighted-strip estimate of Scarlett, Martinez, and Guill\'en i F\`abregas
[\emph{Proc.\ IEEE ISIT}, 2015]. A pairing argument verifies the required covariance nondegeneracy uniformly over all reconstruction words whose distortion balls meet a typical set, yielding the matching converse.
\end{abstract}

 
\section{Introduction} 
{\em Lossy source coding}, a fundamental problem in Shannon theory, seeks to compress a source subject to a prescribed reconstruction distortion. For a discrete memoryless source $X^n=(X_1,\ldots,X_n)\sim P^n$ and a single-letter distortion function $d$, Shannon~\cite{Sha59} showed that the minimum asymptotically achievable compression rate under an average distortion constraint $D$ is the {\em rate-distortion function}
\begin{align}
    R(P,D)
    =\min_{\qyx:\,\bbE_{P\times \qyx}[d(X,Y)]\le D} I(P,\qyx),
\end{align}
where the minimization is over all conditional distributions $\qyx$, or {\em test channels}, from the source alphabet to the reconstruction alphabet. Lossy source coding underpins widely used image and video compression standards, including JPEG and MPEG.

Beyond this first-order characterization, Marton~\cite{Marton74} used the {\em type covering lemma} to determine the {\em error exponent}  which governs the exponential decay of the excess-distortion probability at fixed rates above $R(P,D)$.  For fixed rates below $R(P,D)$, Omura~\cite{Omura75} obtained an early bound on the probability of meeting the distortion constraint. The optimal {\em strong converse exponent}, which characterizes the exponential decay of this probability, is given in Csisz\'ar and K\"orner~\cite[Problem~9.6]{Csi97}.  The present author~\cite{Tan12} studied the {\em moderate deviations} regime.

In this paper, we study {\em fixed-error asymptotics}, where the   excess-distortion probability is constrained to be at most a fixed $\eps\in(0,1)$, independent of the blocklength. Our object of interest is $M^*(n,D,\eps)$, the minimum number of reconstruction codewords required by a length-$n$ code whose probability of exceeding distortion $D$ is at most $\eps$. 

Under their respective regularity assumptions, Ingber and Kochman~\cite[Theorem~1]{ingber11} and Kostina and Verd\'u~\cite[Theorem~12]{kost12} independently established the {\em second-order expansion}
\begin{align}
    \log M^*(n,D,\eps) =nR(P,D)+\sqrt{nV(P,D)}\rvQ^{-1}(\eps)+O(\log n) \label{eqn:second_intro}
\end{align}
for fixed $\eps\in(0,1)$, where $\rvQ$ is the complementary CDF of the standard Gaussian distribution.  The coefficient $V(P,D)$ is called the {\em excess distortion dispersion} in~\cite{ingber11} and the {\em rate-dispersion function} in~\cite{kost12}. For finite-alphabet sources satisfying their smoothness assumptions, Ingber and Kochman~\cite[Eqs.~(7)--(9)]{ingber11} express the dispersion as
\begin{align}
    V(P,D)&=\mathrm{Var}_P\big(R'_P(X,D)\big),\quad \mbox{where}\quad R'_P(x,D) :=
    \left.\frac{\partial}{\partial p(x)}R(p,D)\right|_{p=P}.
\end{align}
Here the coordinate derivatives are defined using a differentiable extension of $R(\fndot,D)$ off the probability simplex. Kostina and Verd\'u~\cite[Theorem~12, Eq.~(83)]{kost12} give the representation
\begin{align}
    V(P,D)=\mathrm{Var}_P\big( \jmath(X,D)\big),
\end{align}
where $\jmath(X,D)$ is the $D$-tilted information (cf.\ Eq.~\eqref{eqn:dtilted}). We use this representation in the present work. Under the conditions of Kostina's
thesis~\cite[Theorem~2.2, Eq.~(2.17)]{Kostina_thesis}, the two variance expressions coincide; also see~\cite{ILZF08}.

\subsection{Our Main Contribution}
Our main contribution is to determine the third-order term in~\eqref{eqn:second_intro} for finite-alphabet discrete memoryless sources at interior distortion levels. Specifically, we establish
\begin{align}
    \log M^*(n,D,\eps)
    =nR(P,D)+\sqrt{nV(P,D)}\rvQ^{-1}(\eps)
      +\frac12\log n+O(1),
    \label{eqn:third_intro}
\end{align}
where the remainder is uniform for $\eps$ in any fixed compact subinterval of $(0,1)$. The result includes the zero-dispersion case
$V(P,D)=0$ and does not require a unique or fully supported
optimal reproduction distribution.

Kostina and Verd\'u~\cite{kost12} previously bounded the $O(\log n)$ remainder; these bounds are reviewed in the first remark following Theorem~\ref{thm:main}. Moulin's two ISIT 2017 papers~\cite{MoulinLower17,MoulinUpper17} stated a converse with a $\frac12\log n$ correction and an achievability bound with an additional $\log\log n$ term. However, both proofs rely on incorrect auxiliary distortion-ball estimates, as explained in the fourth remark following Theorem~\ref{thm:main}. We provide independent achievability and converse proofs establishing~\eqref{eqn:third_intro}.

For achievability when $V(P,D)>0$, we apply the quantitative
delta method of Pinelis and Molzon~\cite{PinelisMolzon2016} to obtain a normal approximation for the empirical distortion-ball exponent. We then average the random coding bound, controlling the excess-distortion probability to obtain the third-order term in the log codebook size.

For the converse when $V(P,D)>0$, we retain the exponential weights associated with tilted information and distortion slack in the reverse change of measure step. The multivariate weighted-strip estimate of Scarlett, Martinez, and Guill\'en i F\`abregas~\cite[Lemma~3]{SMG15} then gives an $n^{-1}$ prefactor by controlling both coordinates jointly. A deterministic argument identifying two inequivalent frequent distortion columns verifies the required covariance nondegeneracy uniformly over all relevant reconstruction words. This uniform estimate yields the matching converse.

\subsection{Other Related Work}

Logarithmic corrections have a long history in lossy source coding. Pilc~\cite{Pilc67} obtained an achievability refinement for the minimum expected distortion at a fixed coding rate, and Zhang, Yang, and Wei~\cite{ZYW97} established the matching asymptotic correction for finite-alphabet memoryless sources under their regularity assumptions. Yang and Zhang~\cite{YZ99} extended redundancy achievability results to abstract alphabets. These works already identify a correction proportional to $\frac{\log n}{2n}$ in expected distortion, with a coefficient determined by the slope of the distortion-rate function. Their criterion differs from the fixed excess-distortion probability considered here: an expected-distortion expansion does not determine the fixed-$\eps$  expansion. More recently, Sriramu and Wagner~\cite[Theorem~III.1]{SW26} determined the logarithmic redundancy coefficient for {\em variable-length} coding of uniform sources with permutation-symmetric distortion measures. Under both almost-sure and expected distortion constraints, they established a redundancy of $\frac{\log n}{2n}+o(\frac{\log n}{n})$ in the minimum expected rate. The present paper considers {\em fixed-length} coding under a fixed excess-distortion probability constraint and determines the expansion up to a bounded remainder.

Classical distortion-ball estimates provide another important connection. Yang and Zhang~\cite[Theorem~3]{YZ99} give product-measure ball bounds with an $n^{-1/2}$ prefactor. Related pointwise expansions and distorted waiting-time asymptotics appear in Kontoyiannis~\cite{K00} and Dembo and Kontoyiannis~\cite{DK99}, respectively.  These results identify the local probability scale underlying the $\frac12\log n$ correction.  Related logarithmic corrections also arise in channel coding and exact channel synthesis. Tomamichel and Tan~\cite{TomamichelTan2013} established a $\frac12\log n+O(1)$ upper bound on the third-order term for discrete memoryless channels with positive $\eps$-dispersion. Under a positive reverse-dispersion condition, this matches Polyanskiy's achievability bound~\cite[Corollary~54]{Pol10}. For exact channel synthesis, Sriramu and
Wagner~\cite[Theorems~1 and~2]{SW24} determined the coefficient of $\log n$ in the minimum expected communication length for full-support nonsingular discrete memoryless channels. 

The weighted-strip estimate of Scarlett, Martinez, and Guill\'en i F\`abregas~\cite[Lemma~3]{SMG15} used in our converse generalizes Polyanskiy's scalar exponentially weighted tail bound~\cite[Lemma~20]{Pol10}. Both estimates originate in finite blocklength channel coding.

\section{Problem Formulation and Main Result}
\subsection{Source Model and Definitions}
Let $X^n=(X_1,\ldots,X_n)\sim P^n$ be a length-$n$ block from a discrete memoryless source with finite alphabet $\calX$, and let $\calY$ be a finite reconstruction alphabet. Here $P^n$ denotes the $n$-fold product distribution of $P$, and $X\sim P$ denotes a generic source symbol. By restricting the source alphabet to $\supp(P):=\{x:P(x)>0\}$, we may assume that $P$ has full support on $\calX$.

For a finite alphabet $\calA$, let $\calP(\calA)$ denote the set of probability mass functions on $\calA$. We use $P$ for the fixed source distribution, $p\in\calP(\calX)$ for a generic source distribution, $q\in\calP(\calY)$ for a reproduction distribution, and  $\qyx(y| x)$ for the conditional probability of reproducing $y$ given source symbol $x$.  The source distribution $p$ and conditional distribution  $\qyx$ induce the joint distribution and output marginal $ 
    (p\times \qyx)(x,y) := p(x)\qyx(y|x)$ and 
    $(p\qyx)(y) := \sum_{x\in\calX}p(x)\qyx(y|x)$.
We identify probability mass functions with their probability vectors. For a positive integer $M$, write $[M]=\{1,\ldots,M\}$. All logarithms are natural, so information is measured in nats.

For a deterministic vector $x^n\in\calX^n$, we denote its {\em type} or {\em empirical distribution} as 
\begin{align}
    \hatP_{x^n}(a) =\frac{1}{n}\sum_{i=1}^n\mathbbm{1}\{ x_i = a\},\qquad a\in\calX. 
\end{align}
Let $\calP_n(\calX)$ be the set of all $n$-types (empirical distributions of length-\(n\) vectors). For $\hatP\in\calP_n(\calX)$, let 
\begin{align}
    \calT_n( \hatP):= \big\{x^n \in\calX^n : \hatP_{x^n}= \hatP \big\}
\end{align}
be the {\em type class} of $\hatP$ \cite{Csi97}.  Types and type classes on $\calY$ are defined analogously, with reproduction types denoted by~$\hatQ$. Norms of finite-dimensional vectors are Euclidean, and neighborhoods of probability distributions are relative to the probability simplex.

The {\em distortion function} $d:\calX\times\calY\to[0,\infty)$ is bounded from above. Define the {\em $n$-letter distortion function} and {\em maximum distortion} as  
\begin{align}
    d_n(x^n,y^n):=\frac{1}{n}\sum_{i=1}^n d(x_i, y_i) \quad\mbox{and}\quad d_{\max}:= \max_{x,y}d(x,y).
\end{align}

\begin{definition}[Non-asymptotic fundamental limits for lossy compression]
An {\em  $(n,M,D, \eps)$-code} for the (fixed-length) lossy compression problem consists of an encoder $f_n:\calX^n\to [M]$ and a  decoder $\varphi_n: [M]\to\calY^n$ such that  the {\em probability of excess distortion} 
\begin{align}
    \Pr\big( d_n(X^n, \varphi_n(f_n(X^n)) ) > D) \le\eps.
\end{align}
    For a positive integer $n$, distortion level $D\ge 0$, and error tolerance $\eps\in (0,1)$, define the non-asymptotic fundamental limit
    \begin{align}
        M^*(n,D,\eps):= \min\big\{ M\in\bbN: \exists\, (n,M,D,\eps)\text{-code} \big\}.
    \end{align}
    If the feasible set is empty, $M^*(n,D,\eps)=+\infty$.
\end{definition}

We now review some known results for this non-asymptotic fundamental limit, putting our main result in context. For a source distribution \(p\), define the {\em average distortion}, {\em mutual information}, {\em rate-distortion function}, {\em minimum achievable distortion}, and {\em zero-rate distortion threshold} as follows:
\begin{align}
    \bbE_{p\times \qyx}[d(X,Y)] &:= \sum_{x,y}p(x) \qyx(y|x)d(x,y),\\ 
    I(p,\qyx)&:=\sum_{x,y}p(x)\qyx(y|x) \log \frac{\qyx(y|x) }{ (p\qyx)(y)  },\\
    R(p,D) &:=\min_{\qyx :  \bbE_{p\times \qyx}[d(X,Y)]\le D}I(p, \qyx) , \label{eqn:rd_func}\\
    D_{\min}(p)&:=\sum_x p(x)\min_{y}d(x,y), \\
    D_{\zr}(p)&:=\min_{y}\sum_x p(x)d(x,y) .
\end{align}

Throughout, we assume that the source-distortion pair $(P,D)$ satisfies $D_{\min}(P)<D<D_{\zr}(P)$. Hence, $R(P,D)>0$. Choose an optimal conditional distribution (also called {\em test channel}) $\qyx^*$ that achieves the minimum in the definition of the rate-distortion function in \eqref{eqn:rd_func}. Let $q_* = P\qyx^*$ be the induced output distribution. Let $\lambda>0$ be a {\em distortion multiplier} at $(P,D)$, meaning that $-\lambda$ is a subgradient of $R(P,\fndot)$ at $D$. Equivalently,
\begin{align}
    R(P,D') \ge R(P,D)-\lambda(D'-D)\qquad\mbox{for every}~ D'\ge D_{\min}(P).   \label{eqn:supp_ineq}
\end{align}
Lemma~\ref{lem:properties} establishes uniqueness of the distortion multiplier at every \(D\in(D_{\min}(P),D_{\mathrm{zr}}(P))\). Consequently, \(R(P,\fndot)\) is differentiable on this interval and \(\lambda=-\partial_D R(P,D)\).   Following Kostina and Verd\'u \cite[Sec.~II-C, Eq.~(28)]{kost12}, for a reproduction distribution $q \in \calP(\calY)$ define the {\em  generalized $D$-tilted information} as
\begin{align}
    \Lambda_q(x,\theta):=-\log\sum_y q(y) \rme^{\theta(D  - d(x,y))},\qquad\theta\ge0. \label{eqn:generalized_tilted}
\end{align}
The {\em $D$-tilted information} is
\begin{align}
     \jmath(x,D):=\Lambda_{q_*}(x,\lambda)=-\log Z_x-\lambda D\quad \mbox{where}\quad Z_x := \sum_y q_*(y)\rme^{- \lambda d(x,y)}. \label{eqn:dtilted}
\end{align}
The {\em rate-dispersion function}~\cite{kost12}  is defined as 
\begin{align}
    V(P,D):= \mathrm{Var}_P\big( \jmath(X,D) \big).
\end{align}

We will also use the notations $\phi(x) := \frac{1}{\sqrt{2\pi}}\rme^{-x^2/2}$, $\Phi(t):=\int_{-\infty}^t\phi(x)\,\rmd x$, and $\rvQ(t):=1-\Phi(t)$ to denote the standard Gaussian probability density function, cumulative distribution function (CDF) and complementary CDF respectively. In the following, positive constants denoted by $C, c$ may change from line to line; they do not depend on $n$ or on the variables over which a bound is stated to be uniform.

\subsection{Main Result and Discussions}
Our main result is stated as follows.
\begin{theorem}[Third-order asymptotics of lossy compression]  \label{thm:main} Under the preceding assumptions, as \(n\to\infty\), uniformly for \(\varepsilon\) in any compact subinterval of \((0,1)\),
\begin{align}
    \log M^*(n,D,\eps)=nR(P,D) + \sqrt{nV(P,D)}\rvQ^{-1}(\eps) + \frac{1}{2}\log n+O(1). 
\end{align}
\end{theorem}
Let \(\Delta_n\) denote the {\em third-order remainder in the expansion of \(\log M^*(n,D,\varepsilon)\)}, i.e., 
\begin{align}
    \Delta_n := \log M^*(n,D,\eps)- \big[nR(P,D) + \sqrt{nV(P,D)}\rvQ^{-1}(\eps)\big].
\end{align}
Theorem \ref{thm:main} says that
\begin{align}
    \Delta_n= \frac{1}{2}\log n+O(1).\label{eqn:halflogn}  
\end{align}
The achievability and converse are proved in Sections~\ref{sec:ach} and~\ref{sec:conv}, respectively.  A few remarks are in order. 
    \begin{enumerate}[leftmargin=*,listparindent=1.5em]
        \item Kostina and Verd\'u~\cite[Theorem~12]{kost12} show for general stationary memoryless sources satisfying regularity conditions   that $\Delta_n$ satisfies, for $n$ sufficiently large,
        \begin{align}
            -\frac{1}{2}\log n-K\le \Delta_n\le C\log n+ \log\log n+K, \label{eqn:kv_upperbound}
        \end{align}
        where $C$ and $K$ are source- and distortion-dependent constants.
        For the biased binary  source  under the Hamming distortion~\cite[Theorem~23]{kost12} with $0<D<\min_xP(x)$,
        \begin{align}
             -K\le \Delta_n\le \frac{1}{2}\log n+ \log\log n+K.  \label{eqn:upperbound_halflogn}
        \end{align}
        For a uniform source on an \(m\)-symbol alphabet under Hamming distortion~\cite[Theorem~27]{kost12}, Kostina and Verd\'u show that \(V(P,D)=0\) and that~\eqref{eqn:halflogn} holds for \(0<D<1-1/m\).  Our result tightens \eqref{eqn:kv_upperbound} for discrete memoryless sources and identifies the third-order term.

        \item  For an i.i.d.\ $\mathcal N(0,\sigma^2)$ source under squared-error distortion with $0<D<\sigma^2$, Kostina and Verd\'u~\cite[Theorem~40]{kost12} show that $V(P,D)=1/2$ and that, for fixed $\eps$ and all sufficiently large $n$, \eqref{eqn:upperbound_halflogn} holds for some constant $K$. Mahmood's recent joint source--channel coding bounds~\cite[Theorems~2 and~4]{mahmood26} give the matching expansion  in~\eqref{eqn:halflogn}   upon setting the channel block size to one bit and all erasure probabilities to zero.
\begin{figure}
    \centering
\begingroup%
\normalfont\normalsize%
\pgfmathsetlengthmacro{\ThirdOrderSetX}{\linewidth/17.6}%
\definecolor{ThirdSetInk}{HTML}{14283A}%
\definecolor{ThirdSetMuted}{HTML}{4C6174}%
\definecolor{ThirdSetBorder}{HTML}{D8E3EB}%
\definecolor{ThirdSetGreen}{HTML}{147664}%
\definecolor{ThirdSetGreenLight}{HTML}{E8F7F1}%
\definecolor{ThirdSetGreenBorder}{HTML}{A3CDBF}%
\definecolor{ThirdSetGray}{HTML}{ABB9C5}%
\definecolor{ThirdSetBlue}{HTML}{255489}%
\definecolor{ThirdSetBlueLight}{HTML}{EDF4FE}%
\definecolor{ThirdSetOrange}{HTML}{9B431C}%
\definecolor{ThirdSetOrangeLight}{HTML}{FFF1E7}%
\begin{tikzpicture}[
  x=\ThirdOrderSetX,y=1cm,
  text=ThirdSetInk,
  text left/.style={anchor=west,inner sep=0pt},
  text right/.style={anchor=east,inner sep=0pt},
  panel/.style={draw=ThirdSetBorder,fill=white,rounded corners=6pt},
  space box/.style={draw=ThirdSetBorder,rounded corners=3pt},
  selected set/.style={draw=ThirdSetGreenBorder,fill=ThirdSetGreenLight},
]
  \path[use as bounding box] (0,0) rectangle (17.6,7.21);
  \draw[panel] (0.50,0.35) rectangle (8.64,6.86);
  \draw[panel] (8.96,0.35) rectangle (17.10,6.86);

  \node[text left,font=\bfseries,text=ThirdSetBlue] at (0.94,6.47)
    {Almost-lossless source coding};
  \node[text left,font=\bfseries,text=ThirdSetOrange] at (9.40,6.47)
    {Lossy source coding};
  \node[text left] at (0.94,6.01) {Count the selected source words};
  \node[text left] at (9.40,6.01) {Sum the probabilities in the ball};

  \draw[space box] (0.94,3.60) rectangle (8.20,5.67);
  \draw[selected set,rounded corners=3pt] (1.13,3.78) rectangle (5.04,5.48);
  \node[text left,text=ThirdSetGreen] at (1.32,5.20) {$\iota_n(x^n)\le b$};
  \node[text right,text=ThirdSetMuted] at (7.99,5.37) {$\mathcal X^n$};
  \foreach \pointX/\pointY in {
    1.52/4.67,2.27/4.67,3.02/4.67,3.77/4.67,4.52/4.67,
    1.52/4.16,2.27/4.16,3.02/4.16,3.77/4.16,4.52/4.16}
    \fill[ThirdSetGreen] (\pointX,\pointY) circle[radius=2.2pt];
  \foreach \pointX/\pointY in {
    5.55/4.67,6.30/4.67,7.05/4.67,7.80/4.67,
    5.55/4.16,6.30/4.16,7.05/4.16,7.80/4.16}
    \fill[ThirdSetGray] (\pointX,\pointY) circle[radius=2.2pt];

  \draw[space box] (9.40,3.60) rectangle (16.66,5.67);
  \draw[selected set] (12.89,4.79) ellipse[x radius=1.82cm,y radius=0.70cm];
  \node[text=ThirdSetGreen,inner sep=0pt] at (12.89,5.24) {$\calB_D(x^n)$};
  \node[text right,text=ThirdSetMuted] at (16.45,5.37) {$\mathcal Y^n$};
  \foreach \pointX/\pointY/\pointRadius in {
    11.50/4.63/2.2,12.14/4.74/3.5,12.71/4.50/1.8,
    13.37/4.72/4.3,14.16/4.55/2.5,13.58/4.39/1.5,12.23/4.37/2.7}
    \fill[ThirdSetGreen] (\pointX,\pointY) circle[radius=\pointRadius pt];
  \foreach \pointX/\pointY/\pointRadius in {
    9.79/5.14/2.2,10.52/5.31/3.3,9.94/4.62/4.4,10.46/4.26/1.8,
    15.24/5.17/3.2,16.22/4.86/2.1,15.67/4.55/4.5,16.23/4.19/2.6,15.14/4.21/1.7}
    \fill[ThirdSetGray] (\pointX,\pointY) circle[radius=\pointRadius pt];
  \node[text left,text=ThirdSetMuted] at (9.60,3.83)
    {Dot area $\propto$ probability mass};

  \node[text left] at (0.94,2.83)
    {$M_{\mathrm{count}}=\big|\{x^n:\iota_n(x^n)\le b\}\big|$};
  \node[text left] at (9.40,3.13)
    {$p_{\mathrm{cov}}=\sum_{y^n\in \calB_D(x^n)}q_*^n(y^n)$};
  \node[text left] at (0.94,2.27)
    {$M_{\mathrm{count}}\asymp n^{-1/2}\mathrm{e}^{b}$};
  \node[text left] at (9.40,2.57)
  {$p_{\mathrm{cov}}  \asymp n^{-1/2}\mathrm{e}^{-a_n(x^n)}$
    \; \; (Lower bound in~\eqref{eqn:ball_lb})};
  \node[text left] at (9.40,2.01)
    {$M_{\mathrm{cov}}\asymp 1/p_{\mathrm{cov}}$ at fixed covering probability $\eta$};

  \fill[ThirdSetBlueLight,rounded corners=5pt] (0.80,0.64) rectangle (8.34,1.43);
  \fill[ThirdSetOrangeLight,rounded corners=5pt] (9.26,0.64) rectangle (16.80,1.43);
  \node[text=ThirdSetBlue] at (4.57,1.035)
    {$\log M_{\mathrm{count}}=b-\frac12\log n+O(1)$};
  \node[text=ThirdSetOrange] at (13.03,1.035)
    {$\log M_{\mathrm{cov}}=a_n(x^n)+\frac12\log n+O(1)$};
\end{tikzpicture}%
\endgroup%
    \caption{The origin of the opposite logarithmic corrections in almost-lossless and lossy source coding.}
    \label{fig:compare}
\end{figure}
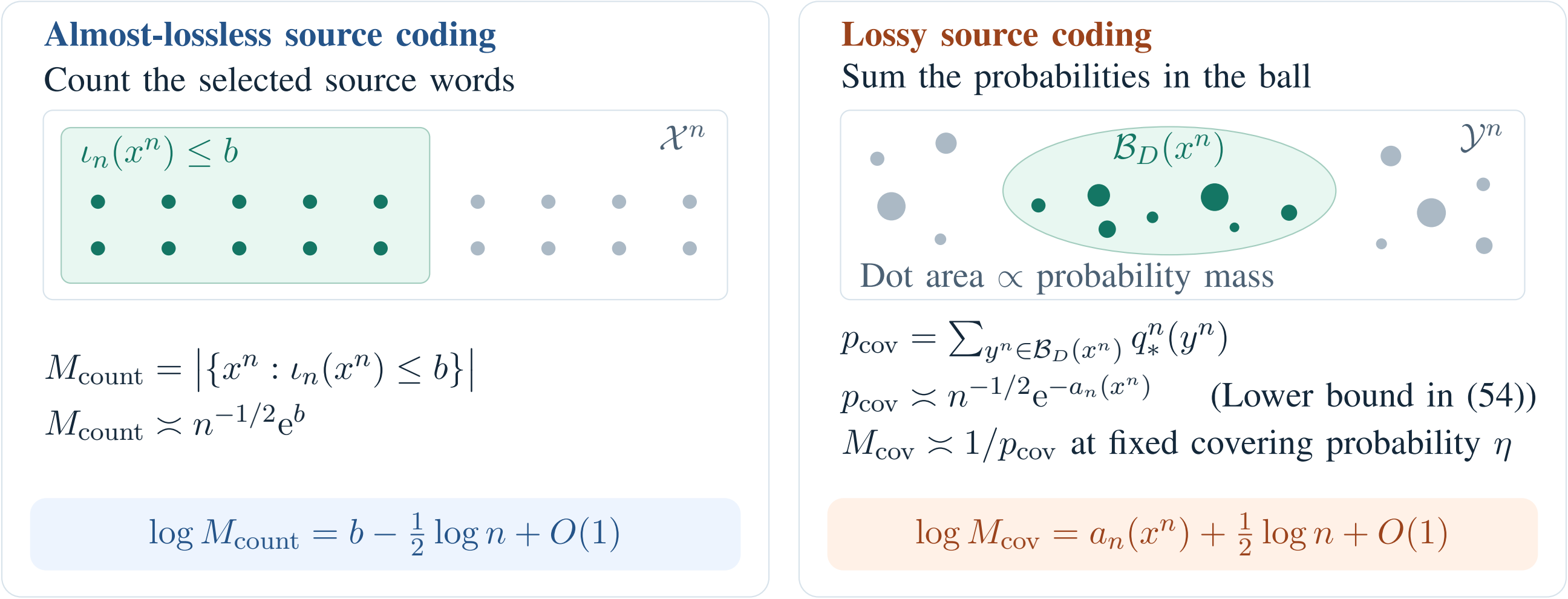
         
        \item Under Hamming distortion, the case $D=0$ corresponds to almost-lossless compression. For a source with positive varentropy $V(P):=\mathrm{Var}_P (-\log P(X))$, Strassen~\cite[Theorem~1.1]{Strassen} showed that, for fixed $\eps$, 
        \begin{align}
            \log M^*(n,0,\eps) =nH(P)+\sqrt{nV(P)}\,\rvQ^{-1}(\eps) -\frac{1}{2}\log n+O(1).
        \end{align}
        The coefficient of $\log n$ changes sign at the lossless endpoint. Fig.~\ref{fig:compare} illustrates how the same $n^{-1/2}$ factor enters the two arguments.

For almost-lossless coding, write $\iota_n(x^n):=-\log P^n(x^n)$, the {\em self-information} of $x^n$. The optimal code represents the most probable source sequences, with ties resolved as needed at a $(1-\eps)$-quantile $b$ of $\iota_n(X^n)$. The number of sequences at or below $b$ is
\begin{align}
    M_{\mathrm{count}}=\big|\{x^n:\iota_n(x^n)\le b\}\big|
    =\rme^b\,\bbE\big[
        \rme^{-(b-\iota_n(X^n))}
        \mathbbm{1}\{\iota_n(X^n)\le b\}
      \big].
\end{align}
The exponential weight makes the count depend mainly on words within a constant distance below $b$. A sufficiently wide window of fixed width there has probability of order $n^{-1/2}$. A window strictly below $b$ also gives a lower bound when only some sequences at $b$ are represented. Thus the code size is of order $n^{-1/2}\rme^b$.

For lossy coding at a fixed interior distortion, fix a source sequence $x^n$ whose type is close to $P$ and draw a reconstruction word from $q_*^n$. Exponentially tilt the reconstruction distribution so that the total distortion has mean $nD$. The change-of-measure weight concentrates the contribution near $nD$, where a window of sufficiently large constant width has tilted probability of order $n^{-1/2}$. The covering probability $p_{\mathrm{cov}}$ is therefore of order $n^{-1/2}\rme^{-a_n(x^n)}$  (cf.\ \eqref{eqn:ball_lb} in Lemma \ref{lem:product_ball} and \cite[Eq.~(3.17)]{YZ99}), where
\begin{align}
    a_n(x^n):=\sup_{\theta\ge 0}\sum_{i=1}^n\Lambda_{q_*}(x_i,\theta).
    \label{eqn:ball_exponent}
\end{align}
Its Taylor approximation about $P$  is 
\begin{align}
 a_n(x^n) = \sum_{i=1}^n\jmath(x_i,D) + O\big(  n \| \hatP_{x^n}-P\|^2\big).     \label{eqn:taylor}
\end{align}
 To cover $x^n$ with probability at least a fixed $\eta\in(0,1)$, the number $M_{\mathrm{cov}}$ of independent reconstruction draws needed is of order $1/p_{\mathrm{cov}}$, hence of order $\sqrt n\,\rme^{a_n(x^n)}$. Taking the logarithm gives the $+\frac{1}{2}\log n$ correction in the achievability argument.
       
        \item We give counterexamples to distortion-ball estimates in Moulin's two ISIT 2017 papers using binary Hamming distortion. Take
\begin{align}
    P(1)=\frac38,\qquad D=\frac14,\qquad q_*(1)=\frac14,
\end{align}
where $q_*$ is an optimal reproduction distribution. Write $R=R(P,D)$ and $\zeta=\log 3$.

In the converse paper~\cite{MoulinLower17}, fix $x^n\in\calT_n(P)$ and let $U_{q_*}$ be the uniform distribution on the type class $\calT_n(q_*)$. For these marginals, $D=1/4$ lies strictly between the minimum distortion $1/8$ and the distortion $7/16$ under independence, as required by Proposition~4.1 in~\cite{MoulinLower17}. The total Hamming distortion is $S:=n d_n(x^n,Y^n)=\frac{5n}{8}-2N_{11}$, where $N_{11}$ counts positions at which both sequences equal one. For $n\equiv8\pmod{16}$, every attainable value of $S$ is odd, whereas $nD$ is even. The shell $\{nD-n^{-1/2}<S\le nD\}$ contains no integer other than $nD$ along this subsequence. Its width in normalized distortion is $n^{-3/2}$, which satisfies the prescribed condition $n^{-2}\ll n^{-3/2}\ll n^{-1}$. This contradicts the positive shell asymptotic in~\cite[Eq.~(20)]{MoulinLower17}.

For $16\mid n$, the boundary $S=nD$ is attained exactly when $N_{11}=3n/16$. Counting and applying Stirling's formula give
\begin{align}
    U_{q_*}\big(S=nD\big)
    =\frac{\binom{5n/8}{n/16}\binom{3n/8}{3n/16}}
            {\binom{n}{n/4}}=\frac{\rme^{-nR}}{\sqrt n}
      \left(\frac{8\sqrt5}{3\sqrt{2\pi}}+o(1)\right).
\end{align}
For marginals $p$ and $q$, define Moulin's fixed-marginal
exponent~\cite[Eq.~(10)]{MoulinLower17} by
\begin{align}
    J(p,q,t)
    :=\min_{\substack{\pi:\,\pi_X=p,\,\pi_Y=q\\
                      \bbE_\pi[d(X,Y)]\le t}}
        \sum_{x,y}\pi(x,y)
        \log\frac{\pi(x,y)}{p(x)q(y)},
\end{align}
where the minimization is over joint distributions $\pi$.
Write $J(t):=J(P,q_*,t)$. For $t$ near $D$, the constraint
is active. Indeed, mixing a joint distribution  of distortion below $t$
with the product distribution  $P\times q_*$, whose distortion
is $7/16$, decreases its mutual information until the
distortion reaches $t$. Let $\pi^{(t)}$ be the unique binary
joint distribution  with marginals $P,q_*$ and Hamming distortion $t$.
Then $J(t)=H(P)+H(q_*)-H(\pi^{(t)})$. The boundary distribution  $\pi^{(D)}$ has entries $(9,1,3,3)/16$ in the order $00,01,10,11$ and attains the rate-distortion minimum, so $J(D)=R$. The four entries of $\pi^{(t)}$ have derivatives $(-1,1,1,-1)/2$ with respect to $t$. Hence
\begin{align}
    \nu=-J'(D)
    =\frac12\log\frac{9\cdot3}{1\cdot3}=\log 3\quad\mbox{and}\quad
    \gamma=J''(D)
    =\frac14\sum_{x,y}\frac{1}{\pi^{(D)}(x,y)}
      =\frac{64}{9}.
\end{align}
The constant in~\cite[Eq.~(19)]{MoulinLower17} is $c_{P,q_*}=4/5$. Consequently, \cite[Eq.~(21)]{MoulinLower17} would imply
\begin{align}
    \limsup_{\substack{n\to\infty\\16\mid n}}
    \sqrt n\,\rme^{nR}U_{q_*} \big(S\le nD\big)
    \le \frac{c_{P,q_*}}{\sqrt{\gamma}\,\nu}
    =\frac{3}{10\log 3}\approx 0.27.
\end{align} The boundary atom alone contributes $8\sqrt5/(3\sqrt{2\pi})\approx 2.38$. This   contradicts the ball bound used to control $\pi_2$ in the  converse~\cite[Eq.~(41)]{MoulinLower17}. There is also a separate variational error in~\cite{MoulinLower17}.
The minimization defining $J(p,q,t)$  in \cite[Eq.~(10)]{MoulinLower17} fixes the
output marginal $q$. In our notation, the Gibbs formula
immediately before  \cite[Eq.~(11)]{MoulinLower17} reads 
    $W^*(y | x)
    =q(y)\exp\{-\nu d(x,y)+\beta(x,\nu)\}$ where $\beta(x,\nu)$ normalizes each row, and $\nu$ is the Lagrange multiplier for the distortion constraint. In general, this formula produces an output marginal different from $q$. Enforcing $pW^*=q$ requires an additional output-dependent Lagrange multiplier in the exponent.

In the achievability paper, \cite[Prop.~5.1, Eq.~(37)]{MoulinUpper17} uses the nonlattice factor $1/\zeta$ for the product reproduction distribution in this example. Under the tilted distribution, the mismatch probabilities at source symbols~$0$ and~$1$ are $1/10$ and $1/2$, respectively. Hence $S$ has mean $nD$, variance $3n/20$, and lattice span one. Exponential tilting and the lattice local central limit theorem give, for $8\mid n$ and every $x^n\in\calT_n(P)$,
\begin{align}
    q_*^n \big(S\le nD\big)
    =\frac{\rme^{-nR}}{\sqrt{2\pi n(3/20)}}
      \left(\frac32+o(1)\right).
\end{align}
The factor $3/2$ comes from summing the contributions of the
layers $S=nD-j$:
\begin{align}
    \sum_{j=0}^{\infty}\rme^{-\zeta j}
    =\frac{1}{1-\rme^{-\zeta}}
    =\frac32
    \ne \frac1{\log 3}.
\end{align}
Eq.~(37)  is used to derive Eq.~(39) in Corollary~5.2 (all in~\cite{MoulinUpper17}), which in turn enters the achievability proof. The converse and achievability derivations thus both require corrected distortion-ball estimates. Theorem \ref{thm:main} confirms the $\frac{1}{2}\log n$ converse coefficient claimed in~\cite{MoulinLower17}  and removes the log $\log n$ term in \cite{MoulinUpper17}. 
     \end{enumerate}
\subsection{Tilted Information and the Reverse Distribution} \label{sec:tilted}
We record several properties of the quantities defined in the previous subsection that aid in the achievability and converse proofs.  For each $y\in\calY$, define the \emph{unnormalized tilted reverse subprobability mass function} $p_y^{(\lambda)}$ and its {\em total mass} $s_y$ by
\begin{align}
    p_y^{(\lambda)}(x) &:=P(x)\rme^{\jmath(x,D)+\lambda(D-d(x,y))}, \label{eqn:ptilt}\\
    s_y&:=\sum_x p_y^{(\lambda)}(x). \label{eqn:sum_ptilt}
\end{align}

\begin{lemma}[Tilted information identities]\label{lem:properties}
At every $D\in(D_{\min}(P),D_{\zr}(P))$, the supporting distortion multiplier is unique and satisfies 
\begin{align}
\lambda=-\partial_D R(P,D)>0.\label{eqn:lambda_partial}     
\end{align}
For every $x\in\calX$, the value of $\jmath(x,D)$ is independent of the choice of optimal test channel $\qyx^*$ (a minimizer in the rate-distortion function \eqref{eqn:rd_func}).

For any optimal test channel $\qyx^*$ with output distribution $q_* = P\qyx^*$,
\begin{align}
    \bbE_{P\times \qyx^*}[d(X,Y)]&=D, \label{eqn:averageD} \\
    \qyx^*(y|x)&=\frac{q_*(y)\rme^{-\lambda d(x,y)}}{Z_x}, \label{eqn:output_test}\\
    \bbE_P[\jmath(X,D)]&=R(P,D). \label{eqn:jmathRPD}
\end{align}
Moreover, $0<s_y\le 1$ for every $y\in\calY$, and $s_y=1$ whenever $q_*(y)>0$.
\end{lemma}
The identities in the lemma are well known in the literature. We include the proof for
completeness.

\begin{proof}
The multiplier is unique if $R(P,\fndot)$ is differentiable at $D$,
since $-\lambda$ is a supporting slope at that point. To apply
Gallager's differentiability result \cite[Corollary~9.4.1]{gallagerIT}, put
\begin{align}
    m(x)&:=\min_y d(x,y),&
    d_0(x,y)&:=d(x,y)-m(x),&
    D_0&:=D-D_{\min}(P).
\end{align}
Replacing $d$ by $d_0$ reduces the mean distortion of every test
channel by $\bbE_P[m(X)]=D_{\min}(P)$. Thus, writing $R_{d_0}$ for the
rate-distortion function under $d_0$, we have $R_{d_0}(P,t)=R(P,t+D_{\min}(P))$.  The distortion $d_0$ has a zero in each row, and $D_0$ lies strictly between zero and its zero-rate distortion threshold. Gallager~\cite[Corollary~9.4.1]{gallagerIT} gives differentiability
at $D_0$, hence at $D$. The unique supporting multiplier is therefore
$\lambda=-\partial_D R(P,D)$. Evaluating the supporting inequality~\eqref{eqn:supp_ineq} at
$D'=D_{\zr}(P)$ gives
\begin{align}
    \lambda\ge \frac{R(P,D)}{D_{\zr}(P)-D}>0.
\end{align}
This proves~\eqref{eqn:lambda_partial}.

Let $\barD$ be the mean distortion of an optimal test channel
$\qyx^*$. Since $\barD\le D$, the inequalities
\begin{align}
    R(P,D)=I(P,\qyx^*)\ge R(P,\barD)
        \ge R(P,D)+\lambda(D-\barD)
\end{align}
force $\barD=D$. This proves~\eqref{eqn:averageD}.

For any test channel $\qyx$ with mean distortion $\tilD$,
the   supporting inequality~\eqref{eqn:supp_ineq} gives
\begin{align}
    I(P,\qyx)+\lambda\tilD
    \ge R(P,\tilD)+\lambda\tilD
    \ge R(P,D)+\lambda D.
\end{align}
By~\eqref{eqn:averageD}, every optimal test channel attains the
right-hand side. Replacing $d$ by $d_0$ shifts the objective by a
constant and thus preserves its minimizers.

For this minimization under $d_0$,
Gallager~\cite[Theorem~9.4.1]{gallagerIT} gives a
unique factor $f_0(x)$ such that every optimal channel satisfies
\begin{align}
    \qyx^*(y|x) =q_*(y)\frac{f_0(x)}{P(x)}\rme^{-\lambda d_0(x,y)}. \label{eqn:channel_rep}
\end{align}
Summing over $y$ gives $f_0(x)=P(x)\rme^{-\lambda m(x)}/Z_x$. Since $f_0(x)$ is unique and $m(x)$ is fixed, both $Z_x$ and $\jmath(x,D)=-\log Z_x-\lambda D$ are independent of the choice of optimal test channel $\qyx^*$. Substituting this expression for $f_0(x)$ into the channel representation  in \eqref{eqn:channel_rep} proves~\eqref{eqn:output_test}.

Using~\eqref{eqn:output_test} and~\eqref{eqn:averageD}, we obtain
\begin{align}
    R(P,D)
    = I(P,\qyx^*)
    = -\bbE_P[\log Z_X]-\lambda D
    = \bbE_P[\jmath(X,D)].
\end{align}
This proves~\eqref{eqn:jmathRPD}.

Finally, substituting the expression for $f_0(x)$ into the
definition of $s_y$ gives
\begin{align}
    s_y=\sum_x f_0(x)\rme^{-\lambda d_0(x,y)}.
\end{align}
Gallager \cite[Eq.~(9.4.9)]{gallagerIT} bounds this sum by one, with equality when $q_*(y)>0$. Finiteness of $d$ and full support of~$P$ give $s_y>0$. Also see~\cite[Eqs.~(15)--(19)]{kost12} for the tilted-information form of these identities.
\end{proof}

\section{Achievability Proof of Theorem \ref{thm:main}} \label{sec:ach}

We use the random coding bound of Kostina and Verd\'u~\cite[Theorem~10]{kost12}, drawing reconstruction words independently from $(q_*)^{n}$. For each source sequence $x^n$ whose type is close to $P$, we lower bound the probability that one reconstruction word covers $x^n$ by a constant times $n^{-1/2}\rme^{-a_n(x^n)}$. The factor $n^{-1/2}$ accounts for the $\frac{1}{2}\log n$ term in the achievable log codebook size.

The main refinement is to retain the random exponent $a_n(X^n)$ when evaluating the covering bound. Kostina and Verd\'u \cite[Lemma~5]{kost12} bound the difference between the empirical exponent and its linear approximation in~\eqref{eqn:taylor} by $C_2\log n$ on an event of probability $1  - O(n^{-1/2})$. This accounts for the extra $C_2\log n$ in $C\log n=(\frac{1}{2}+C_2)\log n$ in~\eqref{eqn:kv_upperbound}. When $V(P,D)>0$, we use~\eqref{eqn:taylor} and the quantitative delta method  of Pinelis and Molzon~\cite[Theorem~2.9]{PinelisMolzon2016} to approximate the distribution of $a_n(X^n)$ directly, with a uniform $O(n^{-1/2})$ error in its distribution function. We then average the conditional covering error directly, which also avoids the separate $\log\log n$ term in \eqref{eqn:kv_upperbound}. When $V(P,D)=0$, a direct bound on the quadratic remainder suffices.

\subsection{Generalized Tilted Information and Distortion Balls}
For $p\in\calP(\calX)$ and $q\in\calP(\calY)$, define 
\begin{align}
    R_0(p,q,D):=\inf_{\qyx :  \bbE_{p\times \qyx}[d(X,Y)]\le D}  \sum_{x,y} p(x)\qyx(y|x) \log\frac{\qyx(y|x) }{q(y) }.
\end{align}
The associated generalized tilted information is $\Lambda_q$ in~\eqref{eqn:generalized_tilted}. Fix an optimal reproduction  distribution $q_*$ and put $Z_x(\theta)=\sum_y q_*(y)\rme^{-\theta d(x,y)}$. The variational representations in \cite[Proposition 1]{DK99} and \cite[Property~1]{YZ99} give the {\em   fixed reproduction distribution exponent}
\begin{align}
    g(p):=R_0(p,q_*,D)  &= \sup_{\theta\ge0}\sum_xp(x)\Lambda_{q_*}(x,\theta) \label{eqn:defR0_former} \\
    &= \sup_{\theta\ge0}\left\{  -\sum_x p(x)\log Z_x(\theta)-\theta D \right\}. \label{eqn:defR0}
\end{align}
The block exponent $a_n(x^n)$ introduced in~\eqref{eqn:ball_exponent} satisfies the identity
\begin{align}
    a_n(x^n)=ng(\hatP_{x^n}).
\end{align}
Thus $g$ is the per-symbol exponent, whereas $a_n$ is the
corresponding block exponent.
Furthermore, in \eqref{eqn:defR0_former}--\eqref{eqn:defR0}, $R_0$ is an optimized expectation of generalized tilted information. Here $q_*$ is held fixed as $p$ varies, while $\theta$ is optimized for each $p$. At the true source distribution $\Lambda_{q_*}(x,\lambda)=\jmath(x,D)$ and $g(P)=R(P,D)$, whereas $g(p)\ge R(p,D)$ in general. The expansion below is the counterpart, with the reproduction distribution  held fixed, of Ingber and Kochman's expansion of the fully optimized $R(p,D)$ \cite[Eq.~(30)]{ingber11}; the corresponding saddlepoint argument appears in \cite[Proof of Theorem 3, Eqs.~(22)--(26)]{DK99}. We  apply Yang and Zhang's product-ball estimate \cite[Theorem~3]{YZ99} uniformly over nearby types.

\begin{lemma}[Uniform product-ball lower bound]\label{lem:product_ball}
There exist a fixed neighborhood $\calU$ of $P$, constants $C,c_0>0$, and an integer $n_0$ such that $g$ is smooth on $\calU$, $g(P)=R(P,D)$, and every $p\in\calU$ satisfies
\begin{align}
    \bigg|g(p)-\sum_xp(x)\jmath(x,D)\bigg|
    \le C\|p-P\|^2. \label{eqn:sec_order}
\end{align}
For every $n\ge n_0$, every $\hatP\in\calP_n(\calX)\cap\calU$, and every $x^n\in\calT_n(\hatP)$, if $Y^n\sim q_*^n$, then
\begin{align}
    q_*^n\big(d_n(x^n,Y^n)\le D\big)
    \ge c_0n^{-1/2}\rme^{-ng(\hatP)}. \label{eqn:ball_lb}
\end{align}
\end{lemma}

\begin{proof}
    Put $F(p,\theta):=\sum_x p(x) \Lambda_{q_*}(x,\theta)$ and $\qyx^{ (\theta) }(y|x)=q_*(y) \rme^{-\theta d(x,y)} /Z_x(\theta)$. The saddlepoint identities \cite[Property~1]{YZ99} give
    \begin{align}
        \partial_\theta F(P,\lambda) :=\frac{\partial F(P,\theta)}{\partial\theta}\Big|_{\theta=\lambda}=0\quad\mbox{and}\quad -\partial_\theta^2 F(P,\lambda)=\sum_x P(x) \mathrm{Var}_{\qyx^*(\cdot|x)}(d(x,Y))>0.
    \end{align}
    The strict positivity follows  because otherwise every row is constant on $\supp(q_*)$, making  $\qyx^*(\cdot|x)=q_*$ and $R(P,D)=0$. The implicit function theorem gives a smooth positive maximizer $\theta(p)$ near $P$ with $\theta(P)=\lambda$. Since $F(p,\fndot)$ is strictly concave, this stationary point is
its unique maximizer. Let 
    \begin{align}
        \alpha(p)=\partial_\theta F(p,\lambda)=\sum_x (p(x)-P(x))\bbE_{\qyx^*(\cdot|x)}[d(x,Y)].
    \end{align}
    After shrinking $\calU$, there is a constant $c>0$ such that $ 
    -\partial_\theta^2F(p,\theta)\ge c$
for every $p\in\calU$ and every $\theta$ between $\lambda$ and~$\theta(p)$. Taylor's theorem gives 
     \begin{align}
     0\le g(p)- \sum_x p(x) \jmath(x,D) &= F(p, \theta(p))-F(p, \lambda) \\
     &\le \alpha(p) (\theta(p)-\lambda)-\frac{c}{2} (\theta(p)-\lambda)^2\\
     &\le \frac{\alpha(p)^2}{2c}\le C\|p-P\|^2.
     \end{align}
     This proves~\eqref{eqn:sec_order}. At $p = P$, this also gives $g(P) = R(P,D)$.

To apply Theorem~3 of Yang and Zhang~\cite{YZ99}, restrict the
reconstruction alphabet to $\supp(q_*)$ and shrink $\calU$ so that
every $p\in\calU$ has full support on $\calX$. Define $\mu_p(\theta)
    :=\bbE_{p\times \qyx^{(\theta)}}[d(X,Y)]$.  The tilted mean satisfies
$\mu_p'(\theta)=-\sum_xp(x) \mathrm{Var}_{\qyx^{(\theta)}(\cdot|x)}[d(x,Y)]<0.$ Every tilted conditional distribution is positive on $\supp(q_*)$,
and at least one distortion row is nonconstant on this support.
Consequently,
\begin{align}
    \sum_xp(x)\min_{y\in\supp(q_*)}d(x,y)
    &<\mu_p(\theta(p))=D<\mu_p(0)=\bbE_{p\times q_*}[d(X,Y)].
\end{align}
These are the distortion-range conditions required by the theorem.
Its finiteness conditions hold because $d$ is bounded.

     Viewing $R_0(p,q_*,t)$ as a function of $t$, the saddlepoint
identities~\cite[Property~3]{YZ99} give
\begin{align}
    \left.\partial_tR_0(p,q_*,t)\right|_{t=D}
    &=-\theta(p),\\
    H_p:=\left.\partial_t^2R_0(p,q_*,t)\right|_{t=D}
    &=\left(
        \sum_xp(x)
        \mathrm{Var}_{\qyx^{(\theta(p))}(\cdot|x)}
            [d(x,Y)]
      \right)^{-1}.
\end{align}
     The third moment term in \cite[Theorem 3]{YZ99} is 
     \begin{align}
         D_3(p)=\sum_{x,y}p(x) q_*(y) d(x,y)^3.
     \end{align}
     Put $B_3=d_{\max}^3$. By the continuity of $\theta(p) $ and the tilted variances, there exist constants $\theta_+$, $H_-$, $H_+$, such that for a sufficiently small $\calU$,
     \begin{align}
         0<\theta(p)\le\theta_+,\quad  0<H_-\le H_p\le H_+\quad \mbox{and}\quad D_3(p)\le B_3. 
     \end{align}
     Choose a fixed $\gamma>0$ such that $\gamma\sqrt{H_-}/(2\sqrt{2\pi})\ge 16H_+^{3/2}B_3+1$. For a source sequence $x^n$ of type $\hatP \in \calU$, apply Yang and Zhang's Eq.~(3.18)  \cite{YZ99} with source distribution $\hatP$. For $n$ large enough such that $\rme^{-\gamma^2 H_+/(2n)}\ge 1/2$, it yields
     \begin{align}
         \frac{q_*^n\big( d_n(x^n,Y^n)\le D \big)}{n^{-1/2}\rme^{-ng(\hatP)}} &\ge\rme^{-\gamma\theta(\hatP)}\left(\frac{\gamma\sqrt{H_{\hatP}}}{\sqrt{2\pi}} \rme^{-\gamma^2 H_{\hatP}/(2n)}- 16 H_{\hatP}^{3/2}D_3(\hatP)\right)\\
         &\ge\rme^{-\gamma\theta_+}\left(\frac{\gamma\sqrt{H_-}}{2\sqrt{2\pi}}  - 16 H_{+}^{3/2}B_3\right)\ge\rme^{-\gamma\theta_+}.
     \end{align}
     This proves \eqref{eqn:ball_lb} with $c_0:=\rme^{-\gamma\theta_+}$.
\end{proof}

\subsection{The Random Coding Bound and Normal Approximation} \label{sec:rc}
The random coding bound of Kostina and Verd\'u~\cite[Theorem~10]{kost12} is quoted in the notation used here. For $x^n\in\calX^n$, define the reproduction ball $\calB_D(x^n):=\{y^n\in\calY^n:d_n(x^n,y^n)\le D\}$.

\begin{lemma}[Kostina--Verd\'u random coding bound] \label{lem:kv}
For every positive integer $M$, there exist an encoder and a decoder
whose excess-distortion probability $\eps_n$ satisfies
\begin{align}
    \eps_n\le
    \inf_{q_n\in\calP(\calY^n)}
    \bbE\Big[\big(1-q_n(\calB_D(X^n))\big)^M\Big].
\end{align}
\end{lemma}
Using $(1-u)^M\le\rme^{-Mu}$ for $u\in[0,1]$ and choosing $q_n=q_*^n$, we obtain the relaxation
\begin{align}
    \eps_n\le\bbE\Big[\big(1-q_*^n(\calB_D(X^n))\big)^M\Big]\le\bbE\Big[\exp\big(-Mq_*^n(\calB_D(X^n))\big)\Big]. \label{eqn:rc_bd}
\end{align}
We evaluate this bound after establishing the normal approximation for its empirical exponent.

Define the {\em (truncated) empirical exponent} by
\begin{align}
    T_n:=
    \begin{cases}
        a_n(X^n)=ng(\hatP_{X^n}),
            &\hatP_{X^n}\in\calU,\\
        nR(P,D),
            &\hatP_{X^n}\notin\calU.
    \end{cases}
\end{align}
We apply the i.i.d.\ quantitative delta method of Pinelis and Molzon~\cite[Theorem~2.9]{PinelisMolzon2016}, explicitly verifying the required local quadratic bound, finite third moment, and positive variance.

\begin{lemma}[Normal approximation for empirical exponent] \label{lem:normal_approx} If $V(P,D)>0$,  then
    \begin{align}
        \sup_{t\in\bbR}\left|\Pr(T_n\le t) - \Phi\left(\frac{t-nR(P,D)}{\sqrt{nV(P,D)}} \right)\right| \le \frac{C}{\sqrt{n}}.
    \end{align}
\end{lemma}
\begin{proof}
 Let $\calH = \{z\in\bbR^{ |\calX| }: \sum_x z(x)=0\}$ and set $U_i :=e_{X_i}-P$, where $e_x$ is the unit vector at $x$. Then,
 \begin{align}
     \bbE[U_i]=0,\quad \bbE [ \|U_i\|^3]<\infty\quad\mbox{and}\quad   \barU_n:=\frac1n\sum_{i=1}^nU_i=\hatP_{X^n}-P 
 \end{align}
 On $\calH$ define $f(z) :=g(P+z)-R(P,D)$ where $P+z\in\calU$, and zero otherwise. The space $\calH$, equipped with the Euclidean norm, is a
finite-dimensional Hilbert space. The function $f$ is Borel
measurable, satisfies $f(0)=0$, and obeys
$T_n=nR(P,D)+nf(\barU_n)$.  Set $L(z) :=\sum_x z(x)\jmath(x,D)$. Lemma~\ref{lem:product_ball} verifies the local quadratic
condition~\cite[Eq.~(2.1)]{PinelisMolzon2016}:
 \begin{align}
     |f(z)-L(z)|\le C\|z\|^2 ~~\mbox{near }0,\quad L(U_i)=\jmath(X_i,D)-R(P,D)\quad \mbox{and}\quad \bbE[L(U_i)^2]=V(P,D)>0.
 \end{align}
 Theorem~2.9 of~\cite{PinelisMolzon2016} now gives
 \begin{align}
     \sup_{t\in\bbR}\left|\Pr(T_n\le t) - \Phi\left(\frac{t-nR(P,D)}{\sqrt{nV(P,D)}} \right)\right| = \sup_{s\in\bbR}\left|\Pr \left(\frac{\sqrt{n}f(\barU_n)}{\sqrt{V(P,D)}}\le s\right) - \Phi (s) \right| \le \frac{C}{\sqrt{n}}
 \end{align}
 as desired.
\end{proof}
 
\subsection{Completion of Achievability Proof} \label{sec:compl_ach}
We now substitute the product-ball estimate (Lemma~\ref{lem:product_ball}) into the   random coding bound (the relaxation of Lemma~\ref{lem:kv} in \eqref{eqn:rc_bd}). 
\begin{proposition}[Achievable code size] \label{prop:ach} For $\eps$ in a fixed compact subset of $(0,1)$, there is a constant $K_+$ such that for all sufficiently large $n$,
    \begin{align}
        \log M^*(n,D,\eps)\le nR(P,D) + \sqrt{nV(P,D)}\rvQ^{-1}(\eps) + \frac{1}{2}\log n+K_+. 
    \end{align}
\end{proposition}
\begin{proof}
    Write $R=R(P,D)$ and $V=V(P,D)$, and let $\eps\in[\eps_0,\eps_1]\subset(0,1)$. Also define the event $\calE_n:=\{\hatP_{X^n}\in\calU\}$ and set $b=\log M-\frac{1}{2}\log n+\log c_0$. Hoeffding's inequality for the empirical distribution gives $\Pr(\calE_n^c)\le \rme^{-c n} $ for sufficiently large $n$. The  relaxed random coding bound in~\eqref{eqn:rc_bd} and Lemma \ref{lem:product_ball} give
    \begin{align}
        \eps_n &\le\bbE\Big[\exp\big(-Mq_*^n(\calB_D(X^n))\big)\Big] \\
        &\le\bbE\Big[\mathbbm{1}_{\calE_n}\exp\big(-M c_0 n^{-1/2}\rme^{-T_n} \big)\Big] + \Pr(\calE_n^c)\\
        &\le \bbE\left[\exp\big( -\rme^{b-T_n} \big) \right]+\rme^{-cn}.
    \end{align}
    
    For the case $V=V(P,D)>0$, define the standard Gumbel probability density
    \begin{align}
        \kappa(u):=\exp(-u-\rme^{-u}),\quad\mbox{which satisfies}\quad  \int_{\bbR}\kappa(u)\,\rmd u=1  \quad\mbox{and}\quad K_0:=\int_{\bbR}|u|\kappa(u)\, \rmd u<\infty. 
    \end{align}
    Since $\int_{-\infty}^v \kappa(u)\,\rmd u=\exp(-\rme^{-v})$, Tonelli's theorem gives
    \begin{align}
        \bbE\left[  \exp\big( -\rme^{b-T_n} \big) \right]=\bbE\bigg[\int_{-\infty}^{T_n-b}\kappa(u)\, \rmd u\bigg]=\int_{\bbR}\kappa(u)\Pr(T_n>b+u)\, \rmd u.
    \end{align}
    Lemma \ref{lem:normal_approx}, followed by the bound $\sup_z\phi(z) =1/\sqrt{2\pi}$, now yields
    \begin{align}
        \eps_n &\le\int_{\bbR}\kappa(u)\rvQ\left( \frac{b+u-nR}{\sqrt{nV}}\right)\, \rmd u + \frac{C}{\sqrt{n}}+ \rme^{-cn} \\
        &\le \rvQ\left(\frac{b-nR}{\sqrt{nV}} \right)+ \frac{K_0}{\sqrt{2\pi nV}}+ \frac{C}{\sqrt{n}}+ \rme^{-cn} \\
        &\le \rvQ\left(\frac{b-nR}{\sqrt{nV}} \right)+ \frac{C_1}{\sqrt{n}}.
    \end{align}
    Take $b\ge nR+\sqrt{nV}\rvQ^{-1}(\eps)+K$. Choose $\eta>0$ so that $\phi( \rvQ^{-1}(\eps)+u)\ge\eta$ for all $\eps\in [\eps_0,\eps_1]$ and $u\in [0,1]$. For fixed $K>0$ and large enough $n$,
    \begin{align}
        \eps_n &\le \rvQ\left( \rvQ^{-1}(\eps)+ \frac{K}{\sqrt{nV}}\right)+ \frac{C_1}{\sqrt{n}} \\
        &=\eps-\int_{0}^{K/\sqrt{nV}}\phi(\rvQ^{-1}(\eps)+u )\,\rmd u+ \frac{C_1}{\sqrt{n}} \\
        &\le \eps+\frac{C_1-\eta K/\sqrt{V}}{\sqrt{n}} \le\eps
    \end{align}
    when $K\ge C_1\sqrt{V}/\eta$.

    For the case $V=V(P,D) =0$, define the event $\calF_L := \{n \|\hatP_{X^n}-P\|^2\le L\}$. Choose $L>0 $ sufficiently large and then $K \ge CL+\log \log (2/\eps_0)$. Since $J_n:=\sum_{i=1}^n\jmath(X_i, D)=nR$ almost surely, Markov's inequality, \eqref{eqn:sec_order} and~\eqref{eqn:rc_bd} give, for all sufficiently large $n$, and $b\ge nR + K$, 
    \begin{align}
        \Pr(\calF_L^c) &\le\frac{1-\|P\|^2}{L}\le\frac{\eps_0}{2} \\
        T_n &\le nR + Cn \| \hatP_{X^n}-P\|^2\le nR+ CL\qquad \mbox{on }\calF_L \subset \calE_n,\\
        \eps_n &\le \exp\big( -\rme^{K-CL}\big)+\frac{\eps_0}{2}\le\eps_0\le\eps.
    \end{align}

    In either case, put $b_\eps:=nR+\sqrt{nV}\rvQ^{-1}(\eps)+K$ with the corresponding $K$, and choose 
    \begin{align}
        M := \big\lceil \rme^{b_\eps+\frac{1}{2}\log n-\log c_0 }\big\rceil.
    \end{align}
    Then $b\ge b_\eps$. Since $R>0$, for all sufficiently large $n$, 
    \begin{align}
        \log M^*(n,D,\eps)\le \log M\le b_\eps+\frac{1}{2}\log n-\log c_0+\log 2.
    \end{align}
    This proves the proposition and completes the proof of the achievability part of Theorem \ref{thm:main}.
\end{proof}

\section{Converse Proof of Theorem \ref{thm:main}} \label{sec:conv}
In the converse proof, we fix a threshold $b$. We bound the probability that a source sequence $X^n$ with $J_n = \sum_{i=1}^n\jmath(X_i,D)\ge b$ is covered by the codebook. For $V=V(P,D)>0$, the key strengthening of Kostina and Verd\'u's converse is a bound of $C\rme^{-b}/n$ on the contribution of each distortion ball, uniformly over its center, outside a common exponentially small  set of source sequences.

To obtain this bound, we retain both exponential weights in the reverse change of measure and apply the multivariate weighted-strip estimate of Scarlett, Martinez, and Guill\'en i F\`abregas \cite[Lemma~3]{SMG15}. This estimate controls the joint concentration of tilted information and distortion, producing an $n^{-1}$ factor.  

We choose $b=\log M-\frac12\log n$ so that the contribution
of the $M$ distortion balls to the union bound is of order
$n^{-1/2}$, i.e., 
$ M {C\rme^{-b}}/{n}=C/{\sqrt n}. $
This threshold $b$ is lower by $\log n$ than the threshold $\log M+\frac12\log n$ used to evaluate Kostina and Verd\'u's converse~\cite[Theorem~7]{kost12}. The normal approximation then improves the third-order lower bound from $-\frac12\log n$ to $+\frac12\log n$. The following subsections establish the uniform ball estimate and complete this argument.

\subsection{Reconstruction Codewords That Cover Typical Inputs} \label{sec:recon}
For a reconstruction codeword $y^n\in\calY^n$, define its source-space distortion ball by
\begin{align}
    \calB_D(y^n):=\big\{x^n\in\calX^n:d_n(x^n,y^n)\le D\big\}. \label{eqn:source_space_ball}
\end{align}
We show that if $\calB_D(y^n)\cap\calA_n\ne\emptyset$, where $\calA_n$ is the typical set defined in \eqref{eqn:typ_A} below, then $y^n$ contains two reconstruction symbols $y,z\in\calY$, each occurring linearly often, such that $d(x,y)-d(x,z)$ is nonconstant in $x\in\calX$. These symbols will verify the full-rank covariance condition required by the weighted-strip estimate
in~\cite[Lemma~3]{SMG15}.

Put $\Delta:=(D_{\zr}(P)-D)/2>0$ and 
\begin{align}
    \calA_n:=\big\{x^n: D_{\zr}(\hatP_{x^n})\ge D+\Delta\big\}. \label{eqn:typ_A}
\end{align} 
Write $y\sim z$ if $d(x,y)-d(x,z)$ is constant in $x\in\calX$. This defines an {\em equivalence relation} on $\calY$. The {\em   equivalence class} containing $y$ is $\{z\in\calY:z\sim y\}$.
Put $\Upsilon:=\max_{x,y,z}|d(x,y)-d(x,z)|$.  Fix $0<\delta<1$ with $\delta \Upsilon<\Delta$. Call a symbol in a reconstruction codeword $y^n$ {\em frequent} if it occurs at least $\delta n/|\calY|$ times. 

\begin{lemma}[Distortion columns of a typical ball] \label{lem:dist} There exist constants $C,c>0$ such that $P^n(\calA_n^c)\le C\rme^{-cn}$ for every $n$. If $\calB_D(y^n)\cap\calA_n\ne\emptyset$, then $y^n$ contains two frequent symbols $y,z$ satisfying $y\nsim z$.
\end{lemma}

\begin{proof}
    Put $\nu_y:=\bbE_P[d(X,y)]$. Since 
$\nu_y 
\ge \min_{z\in\mathcal Y}\mathbb E_P[d(X,z)]
=D_{\zr}(P)
=D+2\Delta$, Hoeffding's inequality gives 
    \begin{align}
        P^n(\calA_n^c) &\le\sum_{y \in \calY} \Pr\left(\frac{1}{n}\sum_{i=1}^nd(X_i,y)<D+\Delta \right)\\ 
        &\le\sum_{y \in \calY} \Pr\left(\frac{1}{n}\sum_{i=1}^n\big(d(X_i,y)-\nu_y\big)<-\Delta \right)\\ 
        &\le |\calY|\rme^{-2n\Delta^2/d_{\max}^2}.
    \end{align}
    At least one symbol of $y^n$ is frequent. Suppose that all frequent symbols lie in one equivalence class $\calC$. Choose $y_0\in\calC$ minimizing $d(x,y)$ within $\calC$ at one $x$. Since the column differences are constant, this $y_0$ minimizes at every~$x$. Every letter outside $\calC$ occurs fewer than $\delta n/|\calY|$ times, so $|\{ i:y_i\notin \calC\}| <\delta n$. Thus, for every $x^n\in\calA_n$,
\begin{align}
    d_n(x^n,y^n)
    &\ge \frac1n\sum_{i=1}^n d(x_i,y_0)
       -\frac{\Upsilon}{n}|\{i:y_i\notin\calC\}|\\
    &\ge D_{\zr}(\hatP_{x^n})-\delta\Upsilon\\
    &\ge D+\Delta-\delta \Upsilon>D.
\end{align}
Hence $\calB_D(y^n)\cap\calA_n=\emptyset$, a contradiction.  
\end{proof}

\subsection{Probability of a Distortion Ball Above a Threshold} \label{sec:prob_dist_ball}

For the next proposition, we define the following notation. Recall  the unnormalized tilted reverse subprobability mass function $p_y^{(\lambda)}$ in \eqref{eqn:ptilt} and its mass $s_y$ in \eqref{eqn:sum_ptilt}. Normalize it to obtain  the probability mass function
\begin{align}
    p_y(x) = \frac{p_y^{(\lambda)}(x)}{s_y}.
\end{align}
For fixed $y^n$, define the {\em reverse product distribution}
\begin{align}
    p_{y^n}(x^n) := \prod_{i=1}^np_{y_i}(x_i).
\end{align} 
Also define the {\em  accumulated $D$-tilted information} and the {\em total
distortion}, respectively, by
\begin{align}
     J_n := \sum_{i=1}^n\jmath(X_i,D)\quad\mbox{and}\quad S_n :=\sum_{i=1}^n d(X_i,y_i).
\end{align}
The component  likelihood ratios multiply to give 
\begin{align}
    \frac{P^n(x^n)}{p_{y^n}(x^n)}= \prod_{i=1}^n \frac{P(x_i)s_{y_i}}{p_{y_i}^{(\lambda)}(x_i)} = \bigg(\prod_{i=1}^n s_{y_i} \bigg)\rme^{-J_n- \lambda(nD-S_n)}\le \rme^{-J_n- \lambda(nD-S_n)}, \label{eqn:ll_ratio}
\end{align}
where $J_n $ and $S_n$ are evaluated at $X^n=x^n$.

The likelihood ratio in~\eqref{eqn:ll_ratio} contains the exponential weights $\rme^{-J_n}$ and $\rme^{-\lambda(nD-S_n)}$. When $V(P,D)>0$, we apply the multivariate weighted-strip bound of Scarlett, Martinez, and Guill\'en i F\`abregas~\cite[Lemma~3]{SMG15} to control the distortion-slack weight $\rme^{-\lambda(nD-S_n)}$ while restricting $J_n$ to an interval of fixed width. When $V(P,D)=0$, $J_n=nR(P,D)$ is deterministic, and Polyanskiy's scalar exponentially weighted tail bound~\cite[Lemma~20]{Pol10} suffices. The two inequivalent frequent symbols identified above ensure the required covariance nondegeneracy after pairing coordinates.

The following proposition is the key in obtaining the $+\frac{1}{2}\log n$ third-order term in the converse.
\begin{proposition}[Uniform ball and tilted information bound]  \label{prop:unif_ball} For $k_0=1+\mathbbm{1}\{V(P,D)>0\}$, every $y^n$ and every $b\in\bbR$ satisfy
\begin{align}
    P^n\big( \calA_n,~S_n\le nD,~J_n \ge b\big)\le C \rme^{-b}n^{-k_0/2}. \label{eqn:unif_ball}
\end{align}
\end{proposition}
\begin{proof}
    Let the event in the probability in \eqref{eqn:unif_ball} be $\calJ_b := \{ \calA_n,~S_n\le nD,~J_n \ge b\}$. Define the {\em scaled distortion slack} $U_n:=\lambda(nD-S_n)$. If the ball $\calB_D(y^n)$ has an empty intersection with $\calA_n$, then the event $\calJ_b$ is empty. Otherwise, Lemma \ref{lem:dist} supplies two inequivalent frequent symbols $y,z$ in $y^n$. Form $k$ disjoint pairs of coordinates, each containing one
occurrence of $y$ and one occurrence of $z$, where
\begin{align}
    \frac{\delta n}{|\calY|}\le k
    \le\left\lfloor\frac n2\right\rfloor.
\end{align}  See Fig.~\ref{fig:pair}. Under $p_{y^n}$, the pair contributions to $(U_n,J_n)$ are i.i.d.\ copies of the random vector
    \begin{align}
        \begin{bmatrix}
\lambda(2D- d(X,y)-d(X',z))\\ \jmath(X,D)+\jmath(X',D)  
        \end{bmatrix}
        \quad \mbox{where}\quad (X,X') \sim p_y \times p_z. 
    \end{align}
    Their covariance has rank $k_0$. Indeed, a linear functional constant on the support would give
    \begin{align}
        \alpha \jmath(x,D)+ \gamma d(x,u)=c_u\quad (x\in\calX, u = y,z)\quad \Longrightarrow\quad \gamma (d(x,y)-d(x,z))=c_y-c_z.
    \end{align}
    Inequivalence forces $\gamma=0$, and if $V=V(P,D)>0$, this then forces $\alpha=0$. If $V=0$, the $\jmath$-coordinate is constant and at least one of the two distortion columns is nonconstant, so the remaining scalar variance is positive.

    \begin{figure}
        \centering
\usetikzlibrary{decorations.pathreplacing}%
\begingroup%
\normalfont\normalsize%
\pgfmathsetlengthmacro{\PairSplitX}{\linewidth/17.6}%
\definecolor{PairSplitInk}{HTML}{14283A}%
\definecolor{PairSplitMuted}{HTML}{4C6174}%
\definecolor{PairSplitBorder}{HTML}{C9D5DF}%
\definecolor{PairSplitBlue}{HTML}{255489}%
\definecolor{PairSplitBlueLight}{HTML}{EDF4FE}%
\definecolor{PairSplitGreen}{HTML}{147664}%
\definecolor{PairSplitGreenLight}{HTML}{E8F7F1}%
\definecolor{PairSplitGrayLight}{HTML}{F3F5F7}%
\noindent\begin{tikzpicture}[
  x=\PairSplitX,y=1cm,text=PairSplitInk,
  text left/.style={anchor=west,inner sep=0pt},
  text right/.style={anchor=east,inner sep=0pt},
  cell text/.style={inner sep=1pt},
  brace/.style={decorate,decoration={brace,mirror,amplitude=4pt},
                draw=PairSplitMuted,line width=.6pt},
  flow/.style={->,>=stealth,draw=PairSplitMuted,line width=.6pt}
]
  \path[use as bounding box] (0,0) rectangle (17.6,6.65);
  \node[text left] at (.26,6.38) {After a permutation of coordinates};
  \node[text right,text=PairSplitMuted] at (17.30,6.38)
    {$n=2k+(n-2k)$};

  \node[font=\bfseries] at (5.26,5.91) {Paired coordinates};
  \node[font=\bfseries] at (13.89,5.91) {Remaining coordinates};

  \foreach \leftX/\firstIndex/\secondIndex/\pairLabel in {
    1.45/i_1/j_1/1,4.10/i_2/j_2/2,7.70/i_k/j_k/k}{
    \pgfmathsetmacro{\midX}{\leftX+.95}
    \pgfmathsetmacro{\rightX}{\leftX+1.90}
    \pgfmathsetmacro{\firstX}{\leftX+.475}
    \pgfmathsetmacro{\secondX}{\leftX+1.425}
    \fill[PairSplitBlueLight] (\leftX,3.64) rectangle (\midX,5.15);
    \fill[PairSplitGreenLight] (\midX,3.64) rectangle (\rightX,5.15);
    \draw[PairSplitBorder,rounded corners=2pt]
      (\leftX,3.64) rectangle (\rightX,5.15);
    \draw[PairSplitBorder] (\midX,3.64) -- (\midX,5.15);
    \draw[PairSplitBorder] (\leftX,4.395) -- (\rightX,4.395);
    \node[cell text,text=PairSplitMuted] at (\firstX,5.44) {$\firstIndex$};
    \node[cell text,text=PairSplitMuted] at (\secondX,5.44) {$\secondIndex$};
    \node[cell text,text=PairSplitBlue] at (\firstX,4.775) {$y$};
    \node[cell text,text=PairSplitGreen] at (\secondX,4.775) {$z$};
    \node[cell text,text=PairSplitBlue] at (\firstX,4.015) {$X_{\firstIndex}$};
    \node[cell text,text=PairSplitGreen] at (\secondX,4.015) {$X_{\secondIndex}$};
    \node[cell text,text=PairSplitMuted] at (\midX,3.34) {pair $\pairLabel$};
  }
  \node at (6.83,4.775) {$\cdots$};
  \node at (6.83,4.015) {$\cdots$};
  \node at (.55,4.775) {$y^n$};
  \node at (.55,4.015) {$X^n$};
  \draw[PairSplitBorder,dashed] (10.10,3.43) -- (10.10,5.57);

  \fill[PairSplitGrayLight,rounded corners=2pt]
    (10.63,3.64) rectangle (17.15,5.15);
  \draw[PairSplitBorder,rounded corners=2pt]
    (10.63,3.64) rectangle (17.15,5.15);
  \draw[PairSplitBorder] (10.63,4.395) -- (17.15,4.395);
  \node[text=PairSplitMuted] at (11.55,5.44) {$r_1$};
  \node[text=PairSplitMuted] at (16.10,5.44) {$r_{n-2k}$};
  \node at (11.55,4.775) {$y_{r_1}$};
  \node at (13.65,4.775) {$\cdots$};
  \node at (16.10,4.775) {$y_{r_{n-2k}}$};
  \node at (11.55,4.015) {$X_{r_1}$};
  \node at (13.65,4.015) {$\cdots$};
  \node at (16.10,4.015) {$X_{r_{n-2k}}$};

  \draw[brace] (1.45,3.03) -- (9.60,3.03);
  \node at (5.525,2.67) {$k$ pairs: $2k$ coordinates};
  \draw[brace] (10.63,3.03) -- (17.15,3.03);
  \node at (13.89,2.67) {$n-2k$ coordinates};
  \draw[flow] (5.525,2.42) -- (5.525,2.16);
  \draw[flow] (13.89,2.42) -- (13.89,2.16);
  \node at (5.525,1.94) {$(U_{\mathrm{pair}},J_{\mathrm{pair}})$};
  \node at (13.89,1.94) {$(U_{\mathrm{rem}},J_{\mathrm{rem}})$};
  \node[text=PairSplitMuted] at (5.525,1.40)
    {$k$ i.i.d.\ contributions under $p_{y^n}$};
  \node[text=PairSplitMuted] at (13.89,1.40)
    {fixed given $\mathcal F_{\mathrm{rem}}$};

  \node at (8.8,.89)
    {$(U_n,J_n)=(U_{\mathrm{pair}},J_{\mathrm{pair}})
      +(U_{\mathrm{rem}},J_{\mathrm{rem}})$};
  \node at (8.8,.15)
    {$\displaystyle\frac{\delta n}{|\mathcal Y|}\le k
      \le\left\lfloor\frac n2\right\rfloor,
      \qquad k=\Theta(n)$};
\end{tikzpicture}%
\endgroup%
        \caption{Illustration of the paired and remaining coordinates in the proof of Proposition \ref{prop:unif_ball} }
        \label{fig:pair}
    \end{figure}
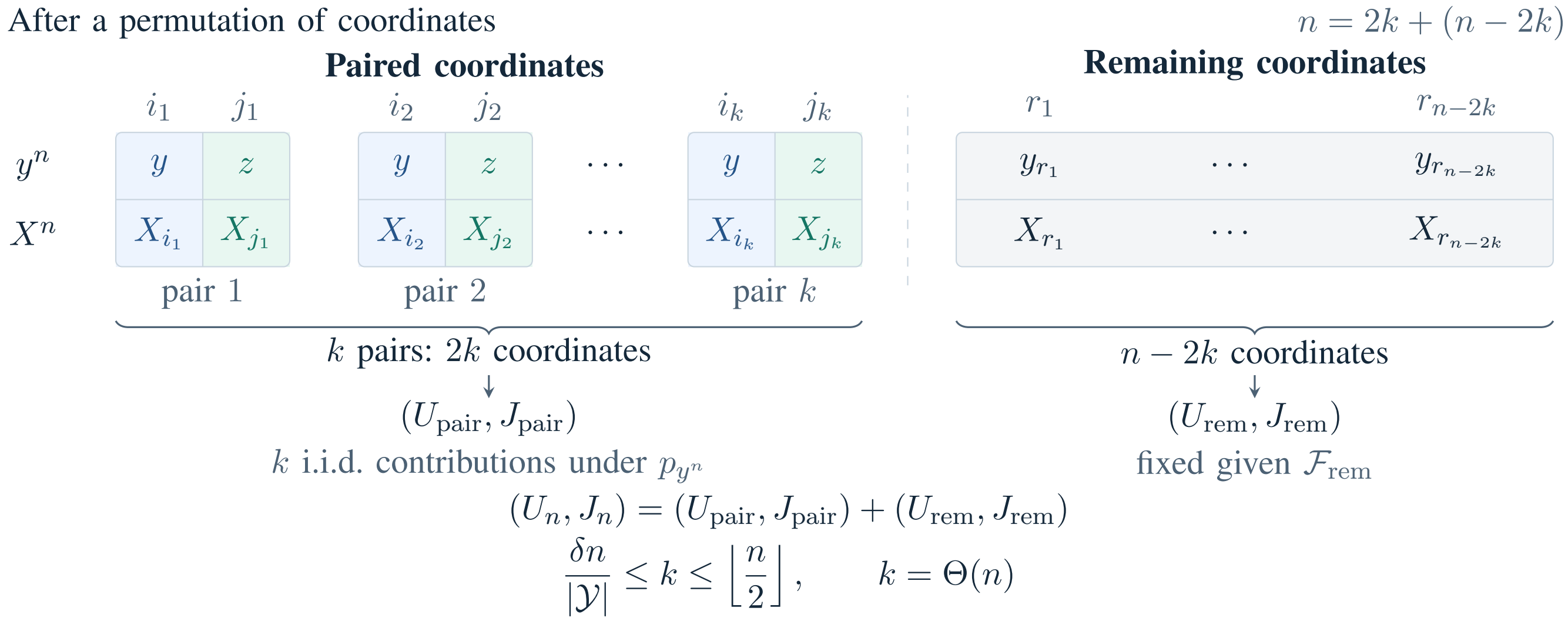

    Split $(U_n, J_n)$ into sums over the paired and remaining coordinates, denoted by subscripts $\mathrm{pair}$ and  $\mathrm{rem}$, and let $\calF_{\mathrm{rem}}$ be the $\sigma$-field generated by the remaining source coordinates.   The exponential-weight estimates in~\cite[Lemma~3]{SMG15} and~\cite[Lemma~20]{Pol10}   give 
    \begin{align}
        &\bbE_{p_{y^n}}\Big[ \rme^{-U_n}\mathbbm{1}\big\{  U_n\ge 0, J_n\in [a,a+1) \big\} \, \Big| \,  \calF_{\mathrm{rem}} \Big]\notag\\
        &=\rme^{-U_{\mathrm{rem}}}
\bbE_{p_{y^n}}\Big[
    \rme^{-U_{\mathrm{pair}}}
    \mathbbm{1}\left\{
        U_{\mathrm{pair}}\ge -U_{\mathrm{rem}},
        J_{\mathrm{pair}}\in
        [a-J_{\mathrm{rem}},a+1-J_{\mathrm{rem}})
    \right\}
    \, \Big| \, \calF_{\mathrm{rem}}
\Big]\\
        &\le \rme^{-U_{\mathrm{rem}}}Ck^{-k_0/2}\rme^{ U_{\mathrm{rem}}}\le C'n^{-k_0/2}.
    \end{align}
    For $k_0=2$, this is~\cite[Lemma~3]{SMG15} with sample size $k$ and strip half-width $1/2$. For $k_0=1$, dropping   the constant $J_n$-strip indicator gives the scalar bound in \cite[Lemma~20]{Pol10}. The scalar bound at the closed threshold follows by applying \cite[Lemma~20]{Pol10} at $-U_{\mathrm{rem}}-h$ and letting $h\downarrow0$. The constants are uniform over the finitely many pairs~$y, z$.

    Average over $\calF_{\mathrm{rem}}$, partition $J_n\ge b$ into unit intervals and use \eqref{eqn:ll_ratio}. For both values of $k_0$,
    \begin{align}
        P^n(\calJ_b)&\le\sum_{\ell\ge 0}\rme^{-b-\ell}\bbE_{p_{y^n}}\Big[ \rme^{-U_n}\mathbbm{1} \big\{  U_n\ge 0, J_n \in [b+\ell, b+\ell+1) \big\} \Big] \\
        &\le C\rme^{-b}n^{-k_0/2}\sum_{\ell\ge0}\rme^{-\ell}=\frac{  C\rme^{-b}n^{-k_0/2}}{1-\rme^{-1}}.
    \end{align}
    This completes the proof of the proposition. 
\end{proof}

\subsection{Completion of Converse}\label{eqn:comp_conv}
For any code, the union of its source-space distortion balls has probability at least its success probability. Intersecting this union with a tilted information tail and applying Proposition \ref{prop:unif_ball} gives the converse immediately.
\begin{proposition} \label{prop:lower_bd_code}
    There exists a constant $K_-\in\bbR$ such that every $(n,M,D,\eps)$-code satisfies
    \begin{align}
        \log M\ge nR(P,D)+\sqrt{nV(P,D)}\rvQ^{-1}(\eps)+\frac{1}{2}\log n-K_-
    \end{align}
    for all sufficiently large $n$, uniformly over $\eps$ in a compact subinterval of $(0,1)$.
\end{proposition}

\begin{proof}
    Given a code with decoder $\varphi_n$, write
$y^n(m):=\varphi_n(m)$ for the reconstruction sequence
associated with index $m\in[M]$. Given the set of reproduction  sequences $\{y^n(1), y^n(2),\ldots, y^n(M)\}$, define its source-space distortion balls as in \eqref{eqn:source_space_ball} and their union by
    \begin{align}
        \calA := \bigcup_{m=1}^M\calB_D(y^n(m)).
    \end{align}
    A source sequence outside $\calA$ cannot be reproduced within distortion $D$. Therefore $P^n(\calA)\ge 1-\eps$. For a threshold $b$, put $\calE:=\calA_n\cap\{J_n\ge b\}$ and $k_0=1+\mathbbm{1}\{V(P,D)>0\}$. Since $P^n(\calA^c)\le\eps$, Proposition~\ref{prop:unif_ball} gives
\begin{align}
    P^n(\calE)-\eps
    &\le P^n(\calA\cap\calE)\\
    &\le M\max_{m\in[M]}
        P^n\big(\calB_D(y^n(m))\cap\calE\big)\\
    &\le CM\rme^{-b}n^{-k_0/2}.
\end{align}
    Since $P^n(\calE)\ge P^n(J_n\ge b)-P^n(\calA_n^c)$, Lemma~\ref{lem:dist} gives
\begin{align}
    \eps
    &\ge P^n(J_n\ge b)
       -CM\rme^{-b}n^{-k_0/2}
       -C\rme^{-cn}.
    \label{eqn:prod_ball_lb}
\end{align}

    For $V=V(P,D)>0$, take $ b=\log M-\frac{1}{2}\log n$. Applying the Berry--Esseen theorem to $J_n$, 
    \begin{align}
        \rvQ\left(\frac{b-nR(P,D)}{\sqrt{nV(P,D)}} \right)&\le P^n(J_n\ge b)+ \frac{C}{\sqrt{n}}\\
        &\le \eps + CM\rme^{-b}n^{-k_0/2} + C\rme^{-cn}+ \frac{C}{\sqrt{n}}\le \eps+\frac{C'}{\sqrt{n}}.
    \end{align}
    The inverse normal tail is Lipschitz on compact subintervals of $(0,1)$. Hence, uniformly for $\eps$ in such an interval and all sufficiently large $n$,
    \begin{align}
        \log M-\frac{1}{2}\log n=b &\ge nR(P,D)+\sqrt{nV(P,D)}\rvQ^{-1} \left(\eps + \frac{C'}{\sqrt{n}}\right)\\
        &\ge nR(P,D)+\sqrt{nV(P,D)}\rvQ^{-1} (\eps )-K_-.
    \end{align}

    For $V=0$, $b=nR(P,D)$ gives $P^n(J_n\ge b)=1$. Then \eqref{eqn:prod_ball_lb} implies that  
    \begin{align}
        M\ge C^{-1}(1-\eps- C\rme^{-cn})\sqrt{n}\rme^{nR(P,D)}.
    \end{align}
    If $\eps\le \eps_{\max}<1$, the factor in brackets is at least $(1-\eps_{\max})/2$ for all sufficiently large $n$, which gives $\log M\ge nR(P,D)+\frac{1}{2}\log n - K_-$.
\end{proof}

Propositions \ref{prop:ach} and~\ref{prop:lower_bd_code} together prove Theorem \ref{thm:main}.

\section{Conclusion and Future Work}

We established the third-order correction $\frac12\log n+O(1)$ for fixed-length lossy compression of finite-alphabet discrete memoryless sources at interior distortion levels. This result complements third-order analyses of channel
coding~\cite{TomamichelTan2013,AW14, TanTom15}, source coding redundancy \cite{SW26}, and channel simulation~\cite{SW24, FSW25,LWK26},  among other studies. The proof combines classical distortion-ball estimates~\cite{ZYW97, YZ99, K00, DK99} with a quantitative normal approximation for achievability and a multivariate weighted-strip estimate for the converse~\cite{SMG15}. Together, these tools yield a bounded remainder uniformly over excess-distortion probabilities in compact subintervals of $(0,1)$.

A natural direction for future work is to investigate refined asymptotics for the minimum covering radius of block codes within the information spectrum framework of Chen and Han~\cite{CH01}. The close connection with rate-distortion theory suggests that the uniform distortion-ball estimates used here may provide a useful starting point. In particular, identifying the higher order terms in the covering radius could clarify how the geometry of distortion balls influences finite blocklength covering limits.

\section*{Acknowledgments}
\emph{Generative-AI use disclosure.}
 The author used OpenAI Codex (GPT-6 Astra, Ultra mode) and Anthropic's Claude (Opus 5) during the development and preparation of this manuscript. These systems assisted in exploring and stress-testing proof strategies, formulating and refining intermediate lemmas, checking algebraic and probabilistic calculations, identifying possible logical gaps, and refining portions of the exposition. The author wrote the manuscript, independently verified every mathematical statement, proof, and reference,  and assumes full responsibility for its contents.
\bibliographystyle{ieeetr} 
\bibliography{isitbib}
\end{document}